\documentclass[%
 letterpaper,
superscriptaddress,
amsmath,amssymb, aps,nobalancelastpage,twocolumn,
pra,
]{revtex4-2}

\usepackage{braket}
\usepackage{graphicx}
\usepackage{dcolumn}
\usepackage{bm}

\usepackage{tikz}
\usetikzlibrary{positioning,chains,quantikz}

\usepackage{adjustbox}
\usepackage{hyperref}

\usepackage{subfigure}
\usepackage{todonotes}
\usepackage{setspace}

\usepackage{soul}

\begin{document}

\preprint{APS/123-QED}

\title{Density Matrix Emulation of Quantum Recurrent Neural Networks for Multivariate Time Series Prediction}

\newcommand{\cesga}{Galicia Supercomputing Center (CESGA), 15705 Santiago de Compostela, Spain}
\newcommand{\cograde}{Computer Graphics and Data Engineering (COGRADE), Departamento de Electrónica e Computación, Universidade de Santiago de Compostela, 15782 Santiago de Compostela, Spain}

\author{J.D. Viqueira}
\email{josedaniel.viqueira@rai.usc.es}
\affiliation{\cesga}
\affiliation{\cograde}
\author{D. Fa\'ilde}
\affiliation{\cesga}
\author{M. M. Juane}
\affiliation{\cesga}
\author{A. G\'omez }
\affiliation{\cesga}
\author{D. Mera}
\affiliation{\cograde}

\date{\today}

\begin{abstract}
Quantum Recurrent Neural Networks (QRNNs) are robust candidates for modelling and predicting future values in multivariate time series. However, the effective implementation of some QRNN models is limited by the need for mid-circuit measurements. Those increase the requirements for quantum hardware, which in the current NISQ era does not allow reliable computations. Emulation arises as the main near-term alternative to explore the potential of QRNNs, but existing quantum emulators are not dedicated to circuits with multiple intermediate measurements. In this context, we design a specific emulation method that relies on density matrix formalism. Using a compact tensor notation, we provide the mathematical formulation of the operator-sum representation involved. This allows us to show how the present and past information from a time series is transmitted through the circuit, and how to reduce the computational cost in every time step of the emulated network. In addition, we derive the analytical gradient and the Hessian of the network outputs with respect to its trainable parameters, which are needed when the outputs have stochastic noise due to hardware errors and a finite number of circuit shots (sampling).
We finally test the presented methods using a hardware-efficient ansatz and four diverse datasets that include univariate and multivariate time series, with and without sampling noise. In addition, we compare the model with other existing quantum and classical approaches. Our results show how QRNNs can be trained with numerical and analytical gradients to make accurate predictions of future values by capturing non-trivial patterns of input series with different complexities.
\end{abstract}

\maketitle


\section{Introduction}

Processing and analysing multivariate time series, generally involve sophisticated algorithms that load high-dimensional datasets evolving in time, wherein temporal correlations are not trivial. In classical computation, Machine Learning is a consolidated approach for prediction and anomaly detection tasks \cite{hochreiter_long_1997, zhang_forecasting_1998, zhang_time_2003, cho_learning_2014, lai_modeling_2018, zhang_deep_2019, blazquez-garcia_review_2021}. Recurrent Neural Networks (RNNs) are a powerful tool for learning sequential data \cite{lipton_critical_2015}, and, over decades, several variations over the first model have improved their performance, like the well-known \textit{Long Short-Term Memory} (LSTM) or the \textit{Gated Recurrent Unit} (GRU) cells \cite{hochreiter_long_1997,cho_learning_2014,cho_properties_2014}. Transformers are a recent alternative  \cite{vaswani_attention_2017} that seems to outperform the previous Neural Network models. However, there is not enough experience when using this algorithm for numerical multivariate time series.

The RNN models feature great results for sequential learning, but they are not problem-free. Some models cannot store information from first inputs in long time series. The LSTM cell arose to address this problem \cite{hochreiter_long_1997}. Moreover, because of the temporal correlations between different variables in multivariate time series, a high-dimensional space appears for computing data with non-linear patterns. Thus, it is necessary to build neural networks with more neurons, layers and parameters, which are computationally expensive and more challenging to train. Rapidly, we have to tackle a Deep Learning task, which requires computational resources and algorithms that ease the parameter optimisation during the neural network training, such as the backpropagation algorithm  for estimating the gradients \cite{schmidhuber_deep_2015}.

Quantum Machine Learning (QML) leverages the power of Quantum Computing to produce complex patterns that are probably not straightforwardly reproducible in a classical computer \cite{dunjko_non-review_2020, biamonte_quantum_2017}. However, in the current \textit{Noisy Intermediate-Scale Quantum} (NISQ) era of quantum computers, there is a need for algorithms requiring 
shallow circuits, since only a limited number of operations are meaningful. Consequently, a substantial part of the research in quantum algorithms is focusing on the Variational Quantum Algorithms (VQAs) \cite{cerezo_variational_2021}. A VQA lies on hybrid classical-quantum routines to classically minimise a cost function computed with a back-end Parameterised Quantum Circuit (PQC).

In this context, a VQA can be used as a Neural Network by feeding a quantum circuit with information from our dataset and then tuning the circuit parameters to minimise the cost function \cite{mitarai_quantum_2018}. The first proposals of Quantum Recurrent Neural Networks (QRNNs) consist of a temporal loop that feeds a quantum state into a quantum circuit and applies the Schrödinger equation to evolve the state. Their applications were simulating stochastic processes \cite{zak_quantum_1999} and stochastic filtering of signals with noise \cite{behera_stochastic_2004, behera_recurrent_2005}. In the middle of the NISQ era and with the explosion of Machine Learning, several proposals have arisen to enhance the power of the current Machine Learning models by adding Parameterised Quantum Circuits as part of the algorithm \cite{chen_quantum_2020,chen_quantum_2022, elkenawy_full-state_2021, chen_quantum_2023, tsai_path_2022}. 

The actual QRNN model seeks to maximise the use and the power of Quantum Computing by a PQC that computes the whole sequence, and the only classical part is data pre-processing, data post-processing and the optimisation of circuit parameters \cite{bausch_recurrent_2020,takaki_learning_2021,siemaszko_rapid_2023,bondarenko_learning_2023,li_quantum_2023,sun_quantum-discrete-map-based_2023}. The most common circuit architecture repeatedly encodes classical data into the quantum circuit, applies a unitary operator and measures a subset of qubits, involving multiple intermediate measurements before the end of the circuit. However, other proposals avoid intermediate measurements at the expense of increasing the number of qubits with the time-series length \cite{bondarenko_learning_2023} or truncating the circuit \cite{sun_quantum-discrete-map-based_2023}.

The QRNN model is expected to provide advantages compared to classical neural networks, which extend to QML models in general. The encoding of classical data into quantum states allows us to compute a number of functions that increases exponentially with the number of qubits \cite{mitarai_quantum_2018}. This feature enhances non-linearities that are required for learning complex patterns. In classical computing, the process requires exponential resources. Besides that, circuits with intermediate measurements, known as dynamic circuits, are not widespread in the QML literature, but they can lead to a new realm of algorithms on real quantum hardware \cite{corcoles_exploiting_2021}, and they are being introduced in Quantum Computing platforms \cite{ibm_bringing_2022,foss-feig_experimental_2023}.

The main aim of this article is to provide the mathematical tools to emulate this type of circuit, which is the core of QRNN algorithms. This is very interesting for two reasons: with the mathematical formulation, we can learn about the propagation of information through the quantum circuit, and the emulation makes the algorithm compatible with classical devices, before sending it to a quantum computer. Within this context, we test the theoretical procedure in several use cases, showing their potential for multivariate time series prediction.

The manuscript is structured as follows.
In Section \ref{sec:information} we provide the method for emulating a quantum circuit with intermediate measurements, which has the property of returning values that depend on past data encoded in the circuit, like recurrent neural networks.
As backpropagation is used in classical Machine Learning, in Section \ref{sec:gradients} we provide a method to analytically compute first- and second-order partial derivatives of the circuit outputs, by an expansion of the Parameter Shift Rule (PSR) \cite{schuld_evaluating_2019,mitarai_quantum_2018}. In Section \ref{sec:results} we show the results after applying the former methods to  an emulation of a QRNN for multivariate time series prediction, in order to demonstrate that the algorithm works by emulating the quantum circuit with the presented method. Finally, discussions are included in Section \ref{sec:conclusions}.

\section{Formulation of the QRNN states with density matrices \label{sec:information}}

The QRNN model is based on the classical RNN model, as proposed by \cite{takaki_learning_2021}.
Following this approach, we explicitly provide a tensor representation of the internal states propagated through the QRNN and its outputs, based on the density matrix formalism. This representation allows the implementation of a classical emulator based on tensor operations and even simple matrix operations, which are very fast in current computational devices.

\subsection{The classical Recurrent Neural Network \label{subsec:crnn}}

Consider a set of data which is time-ordered,
\begin{equation}
    \{ \bm{x}_{(0)}, \bm{x}_{(1)}, \cdots, \bm{x}_{(t)}, \cdots,  \bm{x}_{(T)}\},
\end{equation}
that is the input multivariate time series, since each item is a vector containing the information of $n_v$ variables. The output series is
\begin{equation}
    \{ \bm{y}_{(0)}, \bm{y}_{(1)}, \cdots, \bm{y}_{(t)}, \cdots,  \bm{y}_{(T)}\},
\end{equation}
which is the target in the Machine Learning task.

\begin{figure*}[t]
    \includegraphics[scale=1]{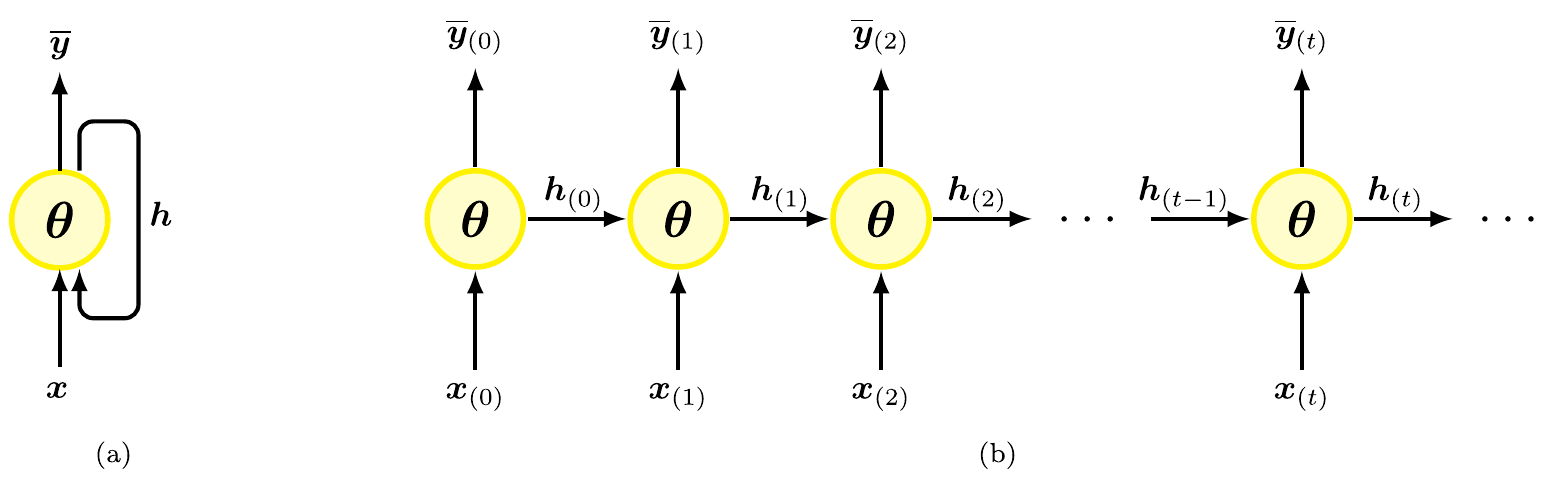}
    \caption{Classical Recurrent Neural Network representations. (a) Basic RNN scheme. (b) Basic RNN unrolled through time.}
\label{fig:rnn}
\end{figure*}

The output of a RNN at time $t$ depends on the inputs from the previous time steps since it preserves past information with a form of memory \cite{geron_hands-machine_2017}. The easiest way to imagine a network with memory is thinking in a box that continuously receives and returns data. To start, the box is supplied with an input, $\bm{x}_{(0)}$, and it returns two objects: an output $\overline{\bm{y}}_{(0)}$ which is read, and a \textit{hidden state} $\bm{h}_{(0)}$ which is re-introduced in the box next time. From the second time onwards, the box receives the previous hidden state $\bm{h}_{(t-1)}$ and the input $\bm{x}_{(t)}$. Then, it computes them and generates the output $\overline{\bm{y}}_{(t)}$ and the hidden state $\bm{h}_{(t)}$ to be re-introduced. The model is represented in Fig. \ref{fig:rnn}, where we can see the recurrence.
The behaviour is dynamic, in contrast to feedforward neural networks, for which the information is never transmitted backwards.

The information flux in a RNN through time splits into three different lines, represented in Fig. \ref{fig:rnn}, that stand for \textit{input data} $\bm{x}_{(t)}$, \textit{output data} $\overline{\bm{y}}_{(t)}$ and \textit{hidden state} $\bm{h}_{(t)}$. Both $\overline{\bm{y}}_{(t)}$ and $\bm{h}_{(t)}$ are functions that depend on $\bm{x}_{(t)}$ and $\bm{h}_{(t-1)}$,
\begin{equation}
    \left\{
    \begin{array}{cc}
         \overline{\bm{y}}_{(t)} & = \mathcal{Y} \left( \bm{x}_{(t)}, \bm{h}_{(t-1)} \right) \\
         \bm{h}_{(t)} & = \mathcal{S} \left( \bm{x}_{(t)}, \bm{h}_{(t-1)} \right) ,
    \end{array}
    \right.
    \label{eq:rnn_funs}
\end{equation}
but at the same time, $\bm{h}_{(t-1)}$ also depends on previous inputs and a hidden state. Functions $\mathcal{Y}$ and $\mathcal{S}$ are the part which is expensive to compute. They consist of matrix calculations that take into account the tunable parameters (weights), structure and connections between the layers of neurons that form the network.

\subsection{The Quantum Recurrent Neural Network \label{subsec:qrnn}}

The QRNN leverages the power of PQCs implicitly performing matrix calculus. Like in the proposed classical model, wherein a network returns output data every time step, we measure the quantum circuit to obtain data in our classical devices. Quantum Networks also contain a \textit{hidden-state} flux $\bm{h}$. As its name suggests, a hidden state propagates internally and carries information not required to be known. That is why we can think of a quantum state as a hidden state. This quantum state is actually a mixed state that arises from the measurement of a subsystem A, and a subsystem B, which is not measured, provided that A and B are entangled.

\begin{figure*}[t]
    \centering
    \includegraphics[scale=1.0]{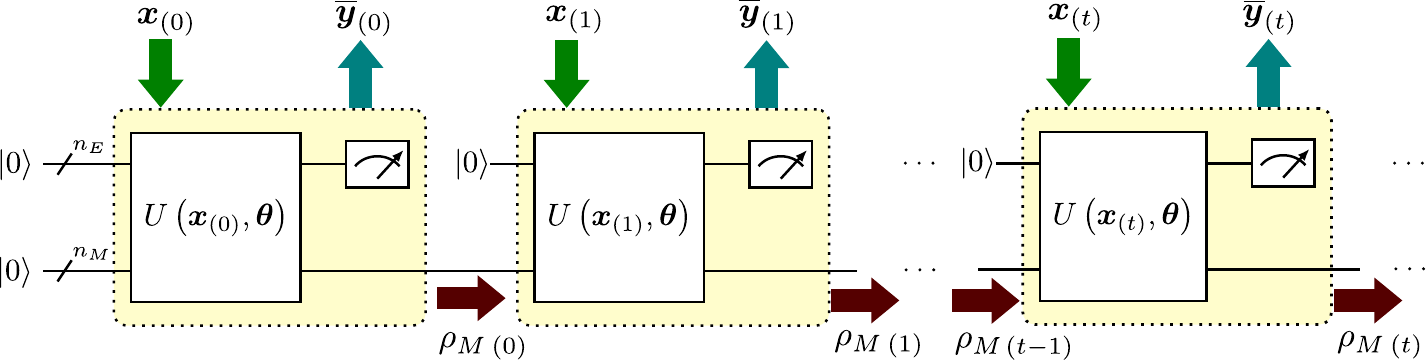}
\caption{General form of the QRNN circuit. Arrows show the information flux.}
\label{fig:qrnn_gen}
\end{figure*}

Hence, we can now construct a general circuit that resembles the classical RNN described above, as it was proposed by \cite{takaki_learning_2021}. Since a quantum circuit diagram represents the series of operations applied over several qubits (vertical direction) during the time (horizontal direction), the circuit in Fig. \ref{fig:qrnn_gen} can be interpreted as the unrolled representation through time of the QRNN. The circuit consists of two quantum registers: an \textit{exchange register} (E) with $n_E$ qubits, and a \textit{memory register} (M) with $n_M$ qubits. The former is used to exchange information between the quantum and the classical interface, by applying encoding and measurement operations, while the latter is never measured. The total number of qubits is $n$. We then have a circuit \textit{block} for every time step: an estimation of an output from the input at time $t$. The instructions are (see Fig. \ref{fig:qrnn_gen}):
\begin{enumerate}
    \item Initialise register M to the $\ket{0}$ state. 
    \item Reset of register E. All register E qubits start at $|0\rangle$ every time step.
    \item Apply a parameterised unitary (ansatz) $U \left( \bm{x}_{(t)}, \bm{\theta} \right)$ that evolves the state of each qubit and entangles some (or all) qubits from both registers E and M, correlating the information from both. $\bm{\theta}$ is the set of variable parameters (weights) of the network.
    \item Measure qubits from register E.
    \item Repeat steps (2), (3), and (4) iteratively from $t=1$ to $t=T$.
\end{enumerate}

The unitary $U \left( \bm{x}_{(t)}, \bm{\theta} \right)$ must encode the classical input $\bm{x}_{(t)}$ each time step, apply parameterised gates depending on the set of weights $\bm{\theta}$, and apply entanglement
between registers E and M. $\bm{\theta}$ is always the same for the different repetitions of the unitary, following the classical approach (see Fig. \ref{fig:rnn}).

Variations of this structure are available in the literature \cite{li_quantum_2023} when trying to create a circuit that sustains the coherence for a longer time. This is not the scope of this work, since we want to establish some bases for designing and emulating quantum RNNs. Solutions for NISQ-era limitations in these networks will require further research that may involve quantum hardware parameters, such as the readout features or the coherence itself  \cite{krantz_quantum_2019}.

\subsection{Memory and output of a QRNN \label{subsec:qrnn_memory_output}}

The way the vanilla RNN \cite{elman_finding_1990} (from which subsequent RNN algorithms were originated) operates data is the ideal sample of how data is explicitly processed in a RNN:
\begin{equation}
    \begin{array}{cl}
        \bm{h}_{(t)} = & \sigma_h \left( W_{hx} \bm{x}_{(t)} + W_{hh} \bm{h}_{(t-1)} + \bm{b}_h \right), \\
        \overline{\bm{y}}_{(t)} = & \sigma_y \left( W_{yh} \bm{h}_{(t)} + \bm{b}_y \right),
    \end{array}
\label{eq:hidden_output}
\end{equation}
being $\{W_{hx}, W_{hh}, W_{yh}\}$ matrices with the weights, $\{\bm{b}_h, \bm{b}_y \}$ the bias vectors, and $\{\sigma_h, \sigma_y \}$ the activation functions.

A density matrix formalism is behind the classical emulation of the quantum circuit in Fig. \ref{fig:qrnn_gen}. The density matrix represents the state of the quantum system as an ensemble of pure quantum states, each of which is described by a state vector. Typically, the state of a quantum circuit evolves unitarily through the application of quantum gates (unitary operators) and can be described by a state vector. However, a single measurement of register E makes the register M collapse to a state $\ket{\psi_s}$, which depends on  the measurement result, $s$, occurring with a probability $p_s$. From a probabilistic perspective, the measurement process gives place to an ensemble of pure states, $\{p_s, \ket{\psi_s} \}$. The density matrix is described as
\begin{equation}
    \rho = \sum_s p_s \ket{\psi_s} \bra{\psi_s} ,
\end{equation}
where $\bra{\psi_s}$ is the hermitian conjugate of $\ket{\psi_s}$.
The density matrix encodes all the information about the ensemble, including the post-measurement state of register M, which is referred to as a \textit{mixed} state.

This formalism provides a powerful method for calculating the exact probability distribution in every measurement of register E qubits.  At the same time, it accounts for non-unitary operations, such as the measurement itself and the reset operation. An alternative is a state vector emulation by sampling outputs, i.e., measurement probabilities are computed from multiple circuit executions, like in a real quantum device.  However, exact probabilities are often preferable when emulating circuits in VQAs, especially when the optimisation method is based on gradients.

Following the instructions to build the quantum circuits, we formulate the expressions to compute two objects for every time step, as in Eq. \eqref{eq:hidden_output}: one is the reduced density matrix $\rho_{M \: (t)}$, which transmits information from $t$ to $t+1$, while the other one is the expected value of some observable $\braket{O}_{(t)}$ (the output before classical post-processing). Following the density matrix formalism, we get
\begin{equation}
    \begin{array}{cl}
         \rho_{M \: (t)} = & \text{Tr}_E \left[ U \: \rho_{(t)} \: U^\dagger \right], \\
         \braket{O}_{(t)} = & \text{Tr } \left[ U \: \rho_{(t)} \: U^\dagger \: O \otimes I^{\otimes n_M} \right],
    \end{array}
\label{eq:rhoM_O}
\end{equation}
where $\rho_{(t)}$ is the initial density matrix of the circuit at time $t$ after the reset on register E and before applying the operator $U = U ( \bm{x}_{(t)}, \bm{\theta} )$. Here, $\text{Tr}_E$ denotes the partial trace over subsystem E and $I$ is the identity operator. $O$ is our observable, which will be, without loss of generality, diagonal. If the observable were not diagonal, we could get its spectral decomposition and apply the necessary transformations to the circuit(s) before measurement. The output (prediction) at time $t$ is
\begin{equation}
    \overline{y}_{(t)} = f ( \braket{O}_{(t)} ),
\end{equation}
being $f$ an arbitrary function. We are restricting to a single-variable output. However, the generalisation for multiple variables is straightforward by considering a set of 
observables instead of a single one.

This section aims to derive a tensor representation of both $\rho_{M \: (t)}$ and $\braket{O}_{(t)}$ to (i) provide an explicit formula that can be implemented in a classical computer as matrix products, and (ii) show how the information is transmitted through the quantum circuit.

For the derivation of the formulas, a symbol and a group of $r$ indices (lowercase letters) represent every mathematical object that is a $r$-rank tensor. The ordering of the indices follows the next criteria. A subindex in parenthesis refers to the current time step, and we can sometimes neglect it. The rest of the indices identify the coefficients for every projector or vector inside the Hilbert space. Indices above (below) correspond to the base vectors in the Hilbert space (dual Hilbert space).  For quantum operators over the full circuit and density matrices representing the $n$ qubits, indices are in pairs. The first index from the pair corresponds with register E, while the second one, with register M. We use the Einstein summation convention, i.e., terms are summed when upper and lower indices are repeated. Note that operations do not commute in general. See Appendix \ref{sec:tensor_notation} for a more comprehensive explanation of the tensor notation used and the operations involved.

The Hilbert space $\mathcal{H}_R$ for each register $R \in \{E,M\}$ has a dimension of $N_R = 2^{n_R}$ and the Hilbert space $\mathcal{H} = \mathcal{H}_E \otimes \mathcal{H}_M$ for the entire system of $n = n_E + n_M$ qubits has a dimension of $N = 2^{n}$.

Tensors describing states at time $t$ and operators are defined in such a way that the following equalities hold. 
\begin{itemize}
    \item \textit{Density matrix} describing the whole system:
    \begin{equation}
        \rho_{(t)} = \sum_{a,b,c,d} \rho^{ab}_{(t) cd} \ket{a \otimes b} \bra{c \otimes d} \in L(\mathcal{H}) .
    \end{equation}
    \item \textit{Reduced density matrix} describing register $M$ state:
    \begin{equation}
        \rho_{M \: (t)} = \sum_{b,d} \rho^{b}_{(t) d} \ket{b} \bra{d} \in L(\mathcal{H}_M) .
    \end{equation}
    \item \textit{State vector} describing register $E$ pure state after reset and $V$:
    \begin{equation}
        \ket{\psi_E}_{(t)} = \sum_{i} \left( \psi_E \right)_{(t)}^i \ket{i} \in \mathcal{H}_E .
    \end{equation}
    \item \textit{Unitary operator} acting over the whole system:
    \begin{equation}
        U_{(t)} = \sum_{a,b,c,d} U^{ab}_{(t) cd} \ket{a \otimes b} \bra{c \otimes d} \in L(\mathcal{H}) .
    \end{equation}
    \item \textit{Kraus operator} acting over the reduced density matrix:
    \begin{equation}
        B_{(t)}^e = \sum_{e,b,d} B^{eb}_{(t)d} \ket{b} \bra{d} \in L(\mathcal{H}_E,\mathcal{H}_M) .
    \end{equation}
    \item \textit{Partial trace} over register E:
    \begin{equation}
    \begin{array}{rcl}
         \text{Tr}_E &: L(\mathcal{H}_E \otimes \mathcal{H}_M) &\rightarrow L(\mathcal{H}_M)  \\
         & \sum_{a,b,c,d} \rho^{ab}_{(t) cd} \ket{a \otimes b} \bra{c \otimes d} &\rightarrow \\
         & \sum_{a,b,c,d} \rho^{ab}_{(t) cd} \ket{b} \bra{d} \delta^c_a &
    \end{array} ,
    \end{equation}
    where $\delta^c_a$ is the Kronecker delta and restricts the sum to the terms having $a=c$.
    \item \textit{Hermitian conjugate}. Represented by $^\dagger$, this operation permutes the above indexes with the below indexes (except $(t)$) and applies the complex conjugate.
\end{itemize}

In the preceding formulas, $L(\mathcal{X},\mathcal{Y})$ denotes the space of bounded linear operators, and $L(\mathcal{X}) = L(\mathcal{X},\mathcal{X})$ \cite{wood2015tensornetworksgraphicalcalculus}. The symbols $\ket{\cdot}$ ($\bra{\cdot}$) represent vectors in the space (dual space), such that
\begin{equation}
    \ket{a} \in \mathcal{H}_E, \: \ket{b} \in \mathcal{H}_M, \: \bra{c} \in \mathcal{H}_E^*, \: \bra{d} \in \mathcal{H}_M^* 
\end{equation}
are basis vectors.
The symbols above (below) can also be interpreted as row (column) indices in matrix form.

The reduced density matrix inmediately before measurement at a time $t$ is given by
\begin{equation}
\begin{split}
     \left( \rho_M \right)^{m}_{(t) n} =
     U^{im}_{(t) kq} \: \rho^{kq}_{(t) lr} \: \left(U^\dagger \right)^{lr}_{(t) jn} \:\delta^{j}_{i} = \\
     \sum_{\bm{i}=0}^{N_E-1} U^{\bm{i}m}_{(t)kq} \: \rho^{kq}_{(t) lr} \: \left(U^\dagger \right)^{lr}_{(t) \bm{i}n} . 
\end{split}
\label{eq:rhoM}
\end{equation}

\begin{figure}[t]
    \begin{adjustbox}{scale=0.80}
    \centering
    \begin{quantikz}
        \ket{0} & \qwbundle{n_E} & 
        \gate[2,label style={rotate=0}]{U \left( \bm{x}_{(t)}, \bm{\theta} \right)} & \qw \\
        & \qwbundle{n_M} & & \qw 
    \end{quantikz}
    =
    \begin{quantikz}
        \ket{0} & \qwbundle{n_E} &
        \gate[1,label style={rotate=0}]{V \left( \bm{x}_{(t)} \right)} &
        \gate[2,label style={rotate=0}]{W \left( \bm{\theta} \right)} & \qw \\
        & \qwbundle{n_M} & \qw & & \qw
    \end{quantikz}
    \end{adjustbox}
    \caption{Decomposition of $U$ operator into encoding ($V$) and evolution part ($W$). The latter is the same for all the circuit blocks since $\bm{\theta}$ does not change during a circuit evaluation.}
    \label{fig:u_decomposition}
\end{figure}

By decomposing the operator into an encoding operator $V(\bm{x}_{(t)})$ that acts only over register E and an operator $W(\bm{\theta})$ that entangles both registers (see Fig. \ref{fig:u_decomposition}), register E is in a pure state $\left( \psi_E \right)_{(t)}^i$ after the application of $V$. The resulting reduced density matrix is computed through an operator-sum representation
\begin{equation}
\begin{split}
    \left( \rho_M \right)^{m}_{(t) n} \\
    = \sum_{\bm{i}=0}^{N_E-1} W^{\bm{i}m}_{(t)kq} \: \left(( \rho_M )^{k}_{(t-1) l} \left( \psi_E^\dagger \right)_{(t) \: r} \left( \psi_E \right)_{(t)}^q  \right) \left(W^\dagger \right)^{lr}_{(t) \bm{i}n} \\
    = \sum_{\bm{i}=0}^{N_E-1} B^{\bm{i}m}_{(t)k} \: ( \rho_M )^{k}_{(t-1) l} \: \left(B^\dagger \right)^{l}_{(t)\bm{i}n},
\end{split}
    \label{eq:rhoM2}
\end{equation}
where
\begin{equation}
    B^{im}_{(t)k} \equiv W^{im}_{kq} \left( \psi_E \right)_{(t)}^q
    \label{eq:kraus}
\end{equation}
are Kraus operators.

Following Eq. \eqref{eq:rhoM_O}, the expectation value of some diagonal observable $O^i_m = d^i \delta^i_m$ (no summing over $i$ since the symbol is above in both terms) at time $t$ is
\begin{equation}
\begin{split}
    \braket{O}_{(t)} = ( \rho'^{kl}_{ij} O^i_m \delta^j_n ) \delta^{mn}_{kl} \\
    = ( \rho'^{kl}_{ij} d^i \delta^i_m \delta^j_n ) \delta^{mn}_{kl} = ( \rho'^{kl}_{ij} d^i \delta^j_n ) \delta^{in}_{kl}
\end{split}
\end{equation}
where $\rho'$ is the density matrix after applying the $U$ operator.

By decomposing this density matrix, we have
\begin{equation}
\begin{split}
    \braket{O}_{(t)} = \sum_{\bm{i}=0}^{N_E-1} d^{\bm{i}} \sum_{\bm{n}=0}^{N_M-1} B^{\bm{n}}_{\bm{i}k} \: ( \rho_M )^{k}_{(t-1) l} \: \left(B^\dagger \right)^{\bm{i}l}_{\bm{n}},
\end{split}
\label{eq:observable_tensor_decomposed}
\end{equation}
which, again, proves the dependency of this observable with respect to both the inputs $\bm{x}_{(t)}$ and the reduced density matrix from the previous step $\rho_{M \: (t-1)}$, provided that $W$ is an operator entangling E and M. In Eq. \eqref{eq:hidden_output}, it depends on the current hidden state, not the previous one. However, the recursion makes it dependent on $\bm{h}_{(t-1)}$ too. We want to remark that this output value is always a real number, due to the general properties of the density matrices.

\subsection{Emulation computational cost}
In most Quantum Machine Learning problems, the $W$ operator is represented by a highly dense matrix (without null elements) and is difficult to decompose since in general it should be formed by several entanglement layers.
During emulation, we build this operator once as a 4-rank tensor. After that, each time step $t$, $N_E$ different $(B^i)_k^m$ operators are computed with Eq. \eqref{eq:kraus}. Therefore, we operate with these $N_M \times N_M$ matrices, avoiding the $N \times N$ matrix size in case we were using the full $U$ operator. Moreover, Eqs. \eqref{eq:rhoM2} and \eqref{eq:observable_tensor_decomposed} are implemented in a single routine, because the terms in \eqref{eq:observable_tensor_decomposed} sum are selected from those of Eq. \eqref{eq:rhoM2} with $n=m$.

Let us assume that the computational cost scales as the product of open and contracted dimensions of tensors, that is, the amount of values that the tensor indices can acquire in aggregate.

The cost of creating $N_E$ different Kraus operators, $B^e$, after $T$ time steps scales $\mathcal{O}(N_E^2N_M^2 T)$, and the cost of generating all the items in Eq. \eqref{eq:rhoM2} after $T$ time steps scales $\mathcal{O}(N_E N_M^3 T)$. In comparison, the cost of evolving the state of the complete system directly with the operator $U$ would scale $\mathcal{O}(N_E^3 N_M^3 T)$.

The cost of a state vector emulation under unitary evolution scales $N = 2^n$. However, each measurement process gives raise to an ensemble $\{p_s, \ket{\psi_s} \}$ of up to $N_E$ elements. Each state vector $\ket{\psi_s}$ is evolved unitarily before the next measurement, that generates again an ensemble of $N_E$ elements. Then, we could only recover the exact expectation values after computing $(N_E)^T$ different quantum trajectories, which arise from collapsing the state to every possibility after measurement. Another option is the state vector emulation of $N_{shots}$ random quantum trajectories. This method consists of randomly selecting a quantum state $\ket{\psi_s}$ from the ensemble with probability $p_s$, which is the probability of measuring result $s$. Each possible measurement result corresponds to a binary string of $n_E$ bits. The emulation needs to be repeated $N_{shots}$ times. In this case, the outputs are not exact, but they have sampling noise, due to the emulation of a finite number of random trajectories. This sampling noise is fatal for gradients computation when finite differences methods are used. 

As time series often involve several time steps and Hilbert spaces grow exponentially with the number of qubits, the most efficient emulation method is the one that uses the operator-sum representation, for most QRNN configurations when computing exact outputs. Even for cases of interest where sampling noise is acceptable, but the number of shots is never fewer than 100, the cost of the operator-sum representation will be preferable.

\section{Analytical derivatives \label{sec:gradients}}

In Machine Learning, most parameter optimisation algorithms are based on gradients, which require some method to be computed. One is numerical differentiation, which is 
inaccurate, and another one is symbolic differentiation, which is more computationally expensive \cite{baydin_automatic_2017}. Automatic Differentiation, with \textit{backpropagation} as the most-known algorithm, seems to have many advantages.

In QML, as current circuits are prone to errors and the outputs are obtained after running the circuit multiple times (shots), systematic and stochastic errors seriously affect the estimation of the cost functions. Errors are amplified when computing numerical gradients, which makes it much more difficult to compute them unless we run circuits with a great number of shots and apply sophisticated error mitigation techniques. Automatic Differentiation as is done in classical computing is not possible, because we need to store intermediate results through the network and intermediate quantum states cannot be accessed in real quantum devices without affecting them. To access them, we need to measure them, and they collapse.

Despite these restrictions, the Parameter Shift Rule (PSR) \cite{mitarai_quantum_2018,schuld_evaluating_2019} provides a method to compute analytical gradients in quantum hardware. Hereafter, we  consider circuits with parameters encoded into rotation gates $R_i(\alpha) = e^{-i \alpha \sigma_i}$ generated by a Pauli matrix $\sigma_i$, because they are a set of fundamental gates used for making PQCs in gate-based quantum devices.

\subsection{First-order partial derivatives \label{subsec:psr1}}

We define the shift expectation value at time $t$, $\left. \braket{O}_{(t)}^\chi \right|_{ri}$, as the expectation value at time $t$ after shifting the parameter $\theta_i$ a value $\chi$ at block $t_r\leq t$ in the quantum circuit. We recall that if $t_r>t$, the shift does not affect the output at time $t$.

For the set of gates that we consider, the shifts are $\pm \frac{\pi}{2}$ \cite{schuld_evaluating_2019}. Then, in the remaining, we write only the sign.

The partial derivative of the observable at time $t$, $\braket{O}_{(t)}$, is then
\begin{equation}
    \partial_i \braket{O}_{(t)} \equiv \frac{\partial \braket{O}_{(t)}}{\partial \theta_i} = \sum_{r=0}^t \frac{1}{2} \left. \left( \braket{O}_{(t)}^+  - \braket{O}_{(t)}^- \right) \right|_{ri},
    \label{eq:psr1}
\end{equation}
derived in Appendix \ref{sec:psr}.

In classical RNNs, the chain rule propagates backwards in time, giving rise to \textit{backpropagation through time} (BPTT) \cite{werbos_backpropagation_1990}. This method is necessary for computing gradients in RNNs, as backpropagation is used in deep learning models, but it has some caveats. Firstly, it requires computing plenty of terms because we need to repeat the network the same number of time steps we have. When contributions from the beginning of the series are negligible, a possible solution is truncation methods \cite{jaeger_tutorial_2002}. Secondly, the propagation backwards in time requires several weights-matrix products. Instability manifests as \textit{vanishing} (\textit{exploding}) gradients if the eigenvalues of the weights matrices are lower (higher) than 1 \cite{zhang_dive_2023}. Truncation can partially address this problem \cite{jaeger_tutorial_2002, tallec_unbiasing_2017}.

From Eq. \eqref{eq:psr1}, computing the partial derivative of the observable at time $t$ requires $2t$ function evaluations with this method. Therefore, computing the gradient requires $2t N_\theta$ function evaluations, being $N_\theta$ the number of parameters of our ansatz. The loss function for training depends on the $T$ time steps. We can evaluate the outputs in a single circuit evaluation, i.e. running the circuit a given number of times (shots) but fixing all the parameters. Thus, in general, the number of function evaluations needed to compute the exact gradient of the loss function is $2T N_\theta$. This is worse than the case of circuits without mid-circuit measurements. Previous proposals implement techniques to reduce the number of circuit evaluations \cite{schuld_evaluating_2019,harrow_low-depth_2021,wierichs_general_2022}, but not for this type of intermediate-measurement-based circuits.

Instability problems cannot appear in Quantum Neural Networks in the form of \textit{exploding gradients} because Quantum Circuits naturally implement unitary operations \cite{benedetti_parameterized_2019} before measuring. Unitary operations are a restriction of some classical ML models, precisely to avoid exploding gradients, and there are ways to implement them without losing too much expressivity \cite{wisdom_full-capacity_2016}. In the case of a QRNN, it is straightforward to see that exploding gradients cannot appear, by looking at Eq. \eqref{eq:psr1} since the subtraction of two observables cannot blow up, and the partial derivative only involves the summation of these terms.

The \textit{vanishing gradient} problem does appear in VQAs, where parameterised circuits can suffer from barren plateaus in the optimisation landscape \cite{mcclean_barren_2018,cerezo_cost_2021}. In the case of RNNs, vanishing gradients cause loss of memory: the network stops depending on past inputs at some moment during the training. In the QRNN model, dependencies of past inputs gradually vanish too, because, at time $t$, the network starts in a mixed state, $\rho_{E \: (t)} \otimes \rho_{M \: (t-1)}$, evolved by a unitary operator. The terms from $\rho_{M \: (t-1)}$ are attenuated after these operations and many circuit blocks.

\subsection{Second-order partial derivatives \label{subsec:psr2}}

Some optimisers require the computation of the Hessian matrix, apart from the Jacobian (gradient). Some authors use the PSR to compute the Hessian \cite{huembeli_characterizing_2021, mitarai_methodology_2019} in PQCs. Here, we provide the method for computing second-order partial derivatives in the QRNN, by applying the PSR.

We define the double-shift expectation value at the time $t$ as $\left. \braket{O}_{(t)}^{\chi \lambda} \right|^{ri}_{sj}$, which is the expectation value at time $t$ after shifting the parameter $\theta_i$ in block $t_r$ a value $\chi$ and the parameter $\theta_j$ a value $\lambda$ in the block $t_s$ of the quantum circuit.

The second-order partial derivatives are
\begin{equation}
\begin{split}
    \partial_i \partial_j \braket{O}_{(t)} \equiv \frac{\partial^2 \braket{O}_{(t)}}{\partial \theta_i \partial \theta_j}\\
    = \frac{1}{4} \sum_r^t \sum_s^t \left. \left( \braket{O}^{++}_{(t)} + \braket{O}^{--}_{(t)}\right) \right|^{ri}_{sj}  - \\
    - \frac{1}{4} \sum_r^t \sum_s^t \left. \left( \braket{O}^{+-}_{(t)} + \braket{O}^{-+}_{(t)} \right) \right|^{ri}_{sj} ,
\end{split}
\end{equation}
for $i \neq j$.

Moreover,
\begin{equation}
\begin{split}
    \partial_i^2 \braket{O}_{(t)} \equiv \frac{\partial^2 \braket{O}_{(t)}}{\partial \theta_i^2} \\ = \frac{1}{2} \sum_r^t \sum_s^r \left. \left( \braket{O}^{++}_{(t)} + \braket{O}^{--}_{(t)} \right) \right|^{ri}_{si} - \\
    - \frac{1}{2} \sum_r^t \sum_s^{r-1} \left. \left( \braket{O}_{(t)}^{+-} + \braket{O}_{(t)}^{-+} \right) \right|^{ri}_{si} - \frac{1}{2} t \braket{O}_{(t)}
\end{split}
\end{equation}
for $i = j$. These are derived in Appendix \ref{sec:psr}.

Considering these expressions, the symmetry $\partial_i \partial_j = \partial_j \partial_i$ and that we can recycle the intermediate calculations for the final estimation of the Hessian of the loss function, the total number of function evaluations is $2 T^2 N^2_\theta + 1$ (proof in Appendix \ref{sec:psr}). The symmetries allow us to reduce the computational cost. However, the scaling with time steps and number of parameters is quadratic. Some approximations would be needed to improve this scaling.

Both gradient and Hessian calculations are easy to parallelise. Then, it would be plausible to simultaneously use multiple Quantum Processing Units (QPUs) to perform several circuit evaluations. The problem is the high noise level of the current quantum hardware and different noise models for different QPUs that could complicate the optimisation process. Further investigation is required to see the scope of these architectures.

\section{Results \label{sec:results}}

\begin{figure*}[t]
    \centering
    \includegraphics[scale=1.0]{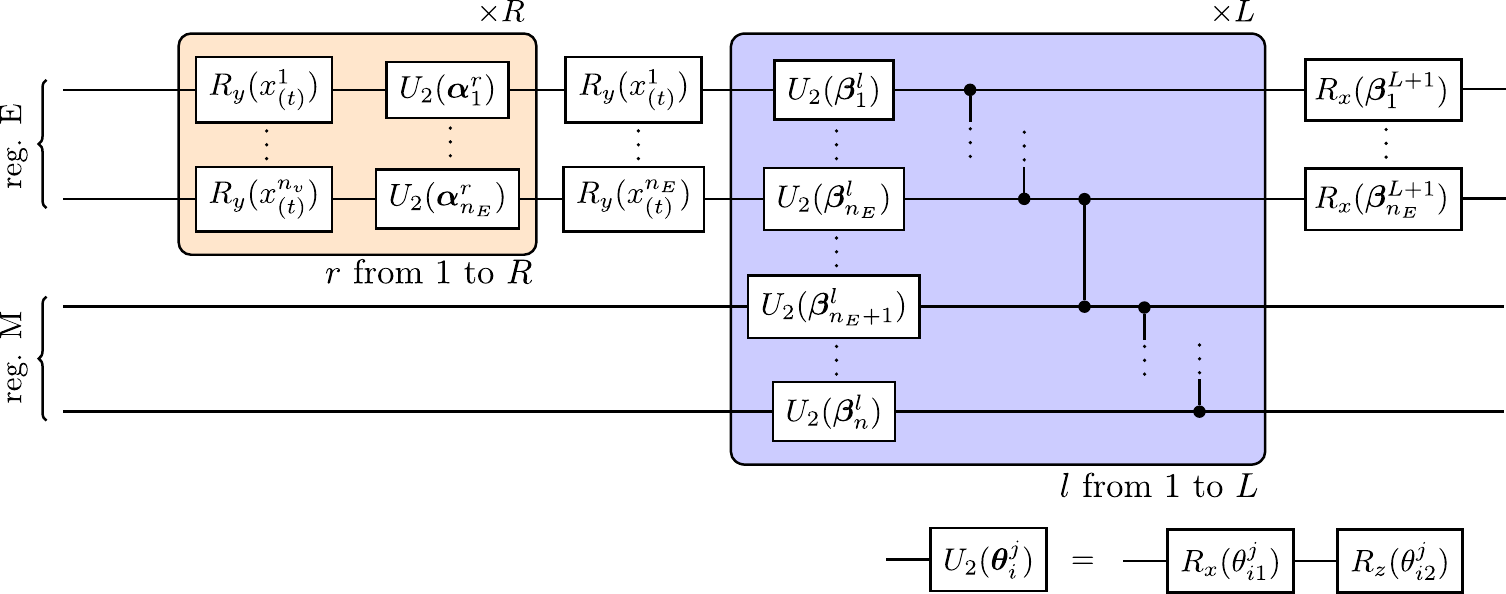}
\caption{QRNN ansatz, $U (\bm{x}_{(t)}, \bm{\theta})$, consisting of two parts. The first one is the data encoding, and gates inside the orange box are repeated with different parameters, that are a subset of trainable parameters, $\bm{\alpha}_i^r \in \{\bm{\theta}\}$. We use one qubit per input variable. The second one is the evolution and entanglement part, where the blue box is repeated $L$ times (layers). Each layer is a column of $R_z R_x$ rotations parameterised by a pair of parameters, $\bm{\beta}_i^l \in \{ \bm{\theta} \}$, and a ladder of CZ gates. A final column of $R_x$ gates is applied over register E before measurement, because the $R_z$ does not affect the $Z$-expectation value.}
\label{fig:qrnn_ansatz}
\end{figure*}

We have implemented an algorithm that emulates a Quantum Recurrent Neural Network, by using the definitions and methods described in the previous sections. The model is based on a quantum circuit that uses a hardware-efficient ansatz. This ansatz uses layers of rotation gates for the encoding of classical inputs into the quantum circuit states, and alternates layers of single-qubit rotation gates, parameterised by the set $\bm{\theta}$, plus ladders of CZ gates.

The circuit ansatz $U(\bm{x}_{(t)},\bm{\theta})$, used to test the performance of the QRNN model, is represented in Fig \ref{fig:qrnn_ansatz}. The encoding part $V(\bm{x}_{(t)}, \bm{\theta})$ acts only over register E qubits and repeats the encoding for each input value $R+1$ times in order to have a better expressivity \cite{schuld_effect_2021,perez-salinas_data_2020,casas_multidimensional_2023}. The evolution and entanglement operator $W(\bm{\theta})$ does not vary during the circuit since $\bm{\theta}$ is fixed inside a circuit evaluation. We aim to maximise its expressibility by repeating the entangling layers several times with different parameters.

In order to test the ideal QRNN emulation and its learning capabilities, we use four datasets: (a) a dimmed triangular signal, (b) a non-linear damping signal with a sinusoidal perturbation, a set (c) consisting of two non-linear damping signals as input and a linear combination of them as output, and (d) the Santa Fe laser series \cite{hubner_dimensions_1989,weigend_results_1993}. (a), (b) and (c) are series of 1000 points, while (d) contains 1980 points. The target is a series with temporal delays with respect to the inputs. These four examples are meaningful for two reasons: the signals are nonlinear, a feature that hinders the training, and are different enough to test several quantum circuit configurations. Moreover, example (c) demonstrates the applicability of the model to, at least, two-variable series, while (d) showcases its effectiveness in predicting a chaotic signal obtained from laboratory, and has been extensively used as a benchmark for time series prediction algorithms. All the datasets were normalized to ensure that values fall within the interval $[-1,1]$. Details about data generation are included in Appendix \ref{sec:details}.

To evaluate the performance of our model against existing algorithms, we have repeated the training with the classical RNN, LSTM and GRU models. Dataset (d)  is employed to assess a Quantum Reservoir computing model described in \cite{mujal_time-series_2023}. This model consists of a non-variational quantum circuit that also incorporates intermediate measurements. We compare their results with ours in the context of predicting the Santa Fe laser series. Three tasks were made with this dataset: prediction with $t_d = \{ 1, 5, 10 \}$, with $t_d$ defined such that $y_{(t)} = x_{(t+t_d)}$, i.e., $t_d$ represents the delay between the input and the target series.

The QRNN is trained by optimising the set of parameters $\bm{\theta}$, using the Adam algorithm implemented in the open-source software framework  \textit{Pennylane} \cite{bergholm2022pennylane}. The \textit{stepsize} parameter is set to 0.001, which is the default in the original algorithm \cite{kingma2017adam}. We divide the series into windows of $T=20$ points and the task is predicting 5 points for the output variable. The windows are divided into three sets: 20 \% for testing, 20 \% of the remaining for validation, and the rest for training. The distribution of validation samples is randomly generated for the three datasets.  The loss function for training is the Mean Squared Error (MSE) between the output $\overline{y}$ and the target series $y$ of each training window (sample). The prediction for every time step is
\begin{equation}
    \overline{y}_{(t)} = \braket{Z^{\otimes n_E}}_{(t)} + b,
\end{equation}
where $Z$ is the $\sigma_z$ Pauli matrix and $b$ is a bias, which is an extra trainable parameter.

The network hyperparameters used for each case are indicated in Table \ref{tab:ml_details}. They were not optimised during this analysis, since the scope of this section is mainly testing the algorithm. However, we systematically increase the number of qubits and layers to have a network with sufficient complexity to make adequate predictions of non-trivial series. Indeed, in (b) we reintroduce data in the second qubit to improve the expressivity. We also want to emphasise that the patterns in (b) and (c) are not periodic. Further work would assess different ansatz architectures, hyperparameters and optimisation techniques.

\begin{table}[t]
\begin{tabular}{c|r|r|r|r|r|l|r|r|}
\cline{2-6} \cline{8-9}
Case & \multicolumn{1}{l|}{$n_E$} & \multicolumn{1}{l|}{$n_M$} & \multicolumn{1}{l|}{$L$} & \multicolumn{1}{l|}{$R$} & \multicolumn{1}{l|}{$N_\theta$} &  & \multicolumn{1}{l|}{$n_{cev}$ num.} & \multicolumn{1}{l|}{$n_{cev}$ ana.} \\ \cline{2-6} \cline{8-9} 
(a)  & 1                          & 2                          & 3                        & 3                        & 26                              &  & 864                                          & 33312                                        \\ \cline{2-6} \cline{8-9} 
(b)  & 2                          & 2                          & 4                        & 1                        & 39                              &  & 1280                                         & 49952                                        \\ \cline{2-6} \cline{8-9} 
(c)  & 2                          & 3                          & 5                        & 3                        & 65                              &  & 2112                                         & 83232                                        \\ \cline{2-6} \cline{8-9} 
(d)  & 1                          & 2                          & 5                        & 3                        & 38                              &  & 2496                                         & 97344                                        \\ \cline{2-6} \cline{8-9} 
\end{tabular}
\caption{Quantum circuit configuration for each case analysed. In case (b), we re-upload input data in two qubits. We add the estimated number of circuit evaluations required per epoch for both numerical and analytical gradient-based parameter optimisations, considering the training size of the datasets and the number of parameters.}
\label{tab:ml_details}
\end{table}

We have executed 10 different parameter optimisations for each dataset, randomly initialising the rotation gates parameters in the interval $[0,2\pi)$. For each epoch, the optimiser runs all the samples in the training dataset, in a random order (shuffle). The result of an optimisation is the set of parameters that return the lowest RMSE in the validation dataset, to avoid overfitting. The result strongly depends on the initialisation. That has consequences in the training, leading to differences in the accuracy of the predictions. Nonetheless, results never show a relative Root MSE (RMSE) greater than 8.5 \% of the signal amplitudes (from -0.75 to 0.75). The optimisation returning the best validation RMSE is selected as the solution among the 10 different optimisation results. After that, we run one optimisation with analytical gradients and no sampling noise, and two groups of 8 optimisations with analytical gradients and sampling noise of $10^3$ and $10^4$ shots, respectively. In order to make a correct comparison, the random seed for parameters initialisation and shuffle are the same as the one from the best numerical optimisation.

\begin{figure*}[t]
    \centering
    \includegraphics[scale=0.65]{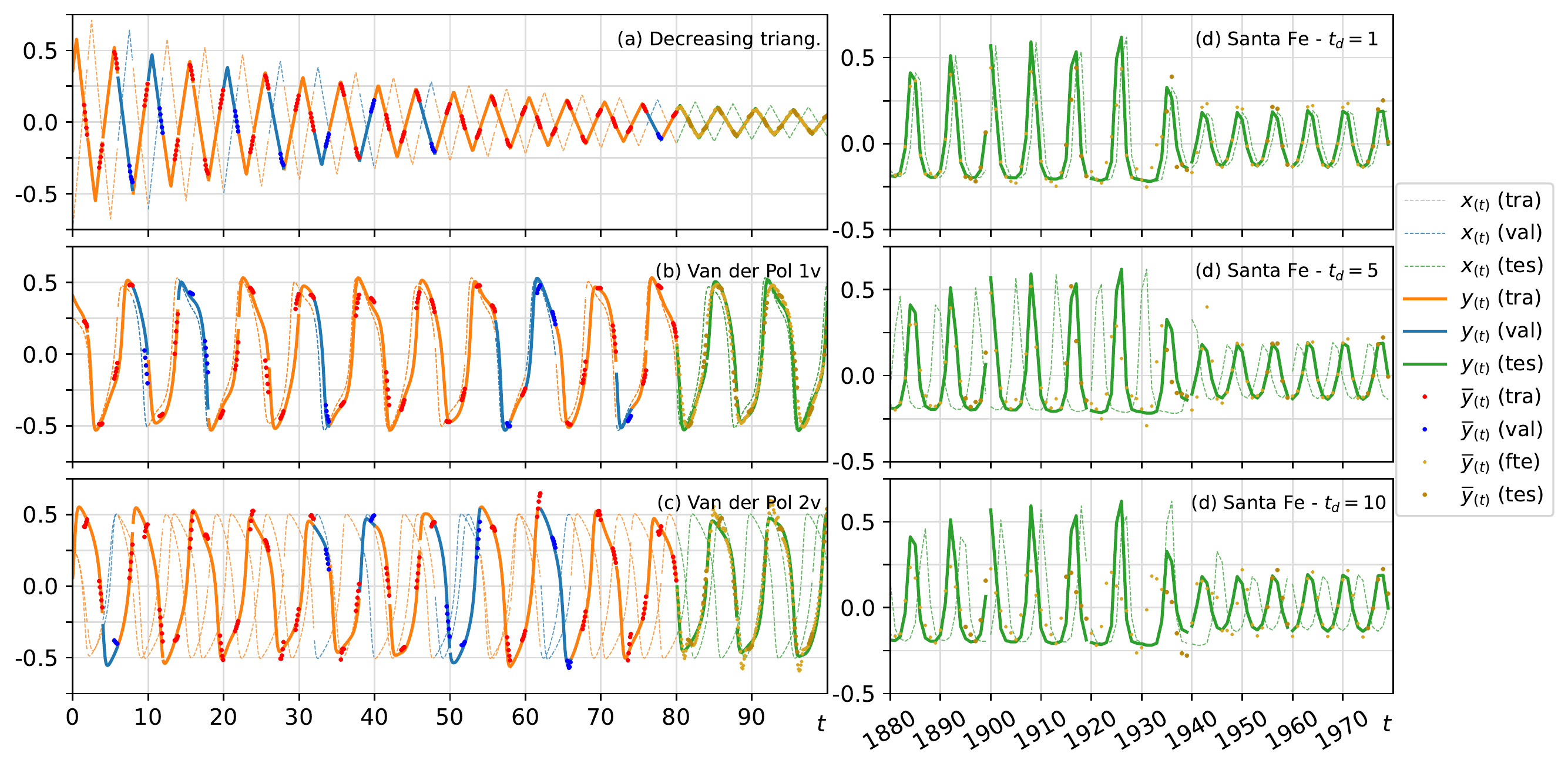}
\caption{Results of the learning task with the QRNN model. Numerical gradients were used for the training, without sampling noise. Each time series was divided into windows of 20 points, each one being a sample. The neural network must predict the value of the last 5 outputs for each window, by reducing the RMSE with respect to the last 5 points of the target. Dashed lines are the inputs $\bm{x}_{(t)}$, from which we make the prediction. Solid lines represent the targets. Points are the predictions. Windows are represented in orange (red), blue (dark blue) or green (gold, dark gold) depending on whether the window is used for training (tra), validation (val) or testing (tes), respectively. In the test region, we add a window to make the prediction for the full test sequence (fte). For the Santa Fe series, a representative portion of the test region is displayed.}
\label{fig:series}
\end{figure*}

\begin{figure*}[t]
    \centering
    \includegraphics[scale=0.45]{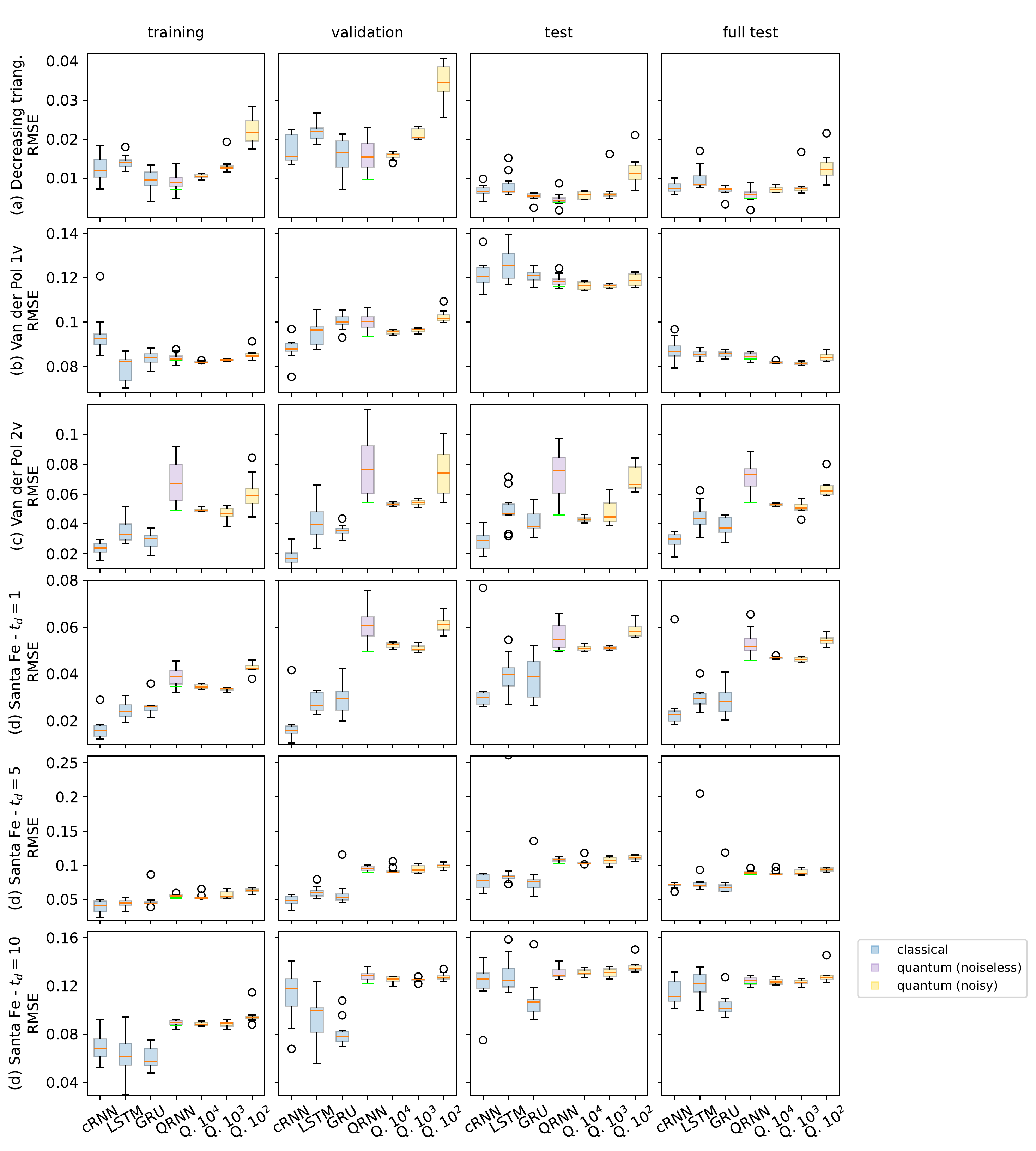}
\caption{Comparison of RMSE results across different models used for predicting each dataset. Box plots illustrate the distributions of RMSE for the classical RNN, LSTM, GRU, and QRNN models, based on 10 realizations with varied random parameter initializations. The box plots for QRNN (``Q. $N_{shots}$'') with finite number of shots represent 8 realisations, all initialised with the same parameter set as the noiseless optimisation giving the lowest validation RMSE. The RMSE values are calculated using outputs from a noiseless emulation of the QRNN, utilizing the parameters from the corresponding training, whether it is noiseless or noisy. The whiskers extend from the box to the farthest data point lying within 1.5 times the inter-quartile range (IQR) from the box. Outliers, i.e., values past the end of the whiskers, are plotted as individual points. The green horizontal line indicates the RMSE value associated with the optimization that resulted in the lowest validation loss.}
\label{fig:results_comparison}
\end{figure*}

The Adam optimiser uses the gradient of the loss function to find its minimum. It is based on adaptive estimations of lower-order moments to change the parameters, allowing stochastic loss functions \cite{kingma2017adam}. This algorithm does not compute the Hessian. We use the forward difference method estimation with an absolute step size $\epsilon  = 1 \times 10^{-7}$, following \textit{Pennylane} defaults, and the analytical form of the gradient.

In ideal emulation without sampling (exact), the emulated quantum circuit returns an expectation value with a precision given by the classical machine variables (e.g. float-64). As, in general, an expectation value cannot be measured from a quantum circuit in a single shot, its evaluation by multiple repetitions (shots) of the circuit leads to stochastic variations. That makes numerical gradients very unstable and we must use analytical gradients. The PSR allows us to tackle this problem in real quantum circuits, but ideal exact emulation using numerical gradients requires fewer circuit evaluations. The density matrix emulation enables exact probabilities calculation, but also simulating stochastic noise to investigate its effect on the algorithm's performance. In the latter case, we evaluate the partial derivatives of the circuit outputs by the formulas provided in Section \ref{sec:gradients}, and then we apply the chain rule to calculate the gradient of the loss function.

For analytical optimisations, we have emulated the QRNN by the provided formulation, and artificially added sampling noise, corresponding to $10^2$, $10^3$ and $10^4$ shots, which are common numbers when evaluating expected values in VQAs. It is simulated with Gaussian noise of standard deviation
\begin{equation}
    \sigma ( \langle Z^{\otimes n_E} \rangle_{(t)} ) = \sqrt{\frac{1 - \langle Z^{\otimes n_E} \rangle_{(t)}^2}{N_\text{shots}}},
\end{equation}
according to the postulates of quantum mechanics.

\begin{figure*}[t]
    \centering
    \includegraphics[scale=0.65]{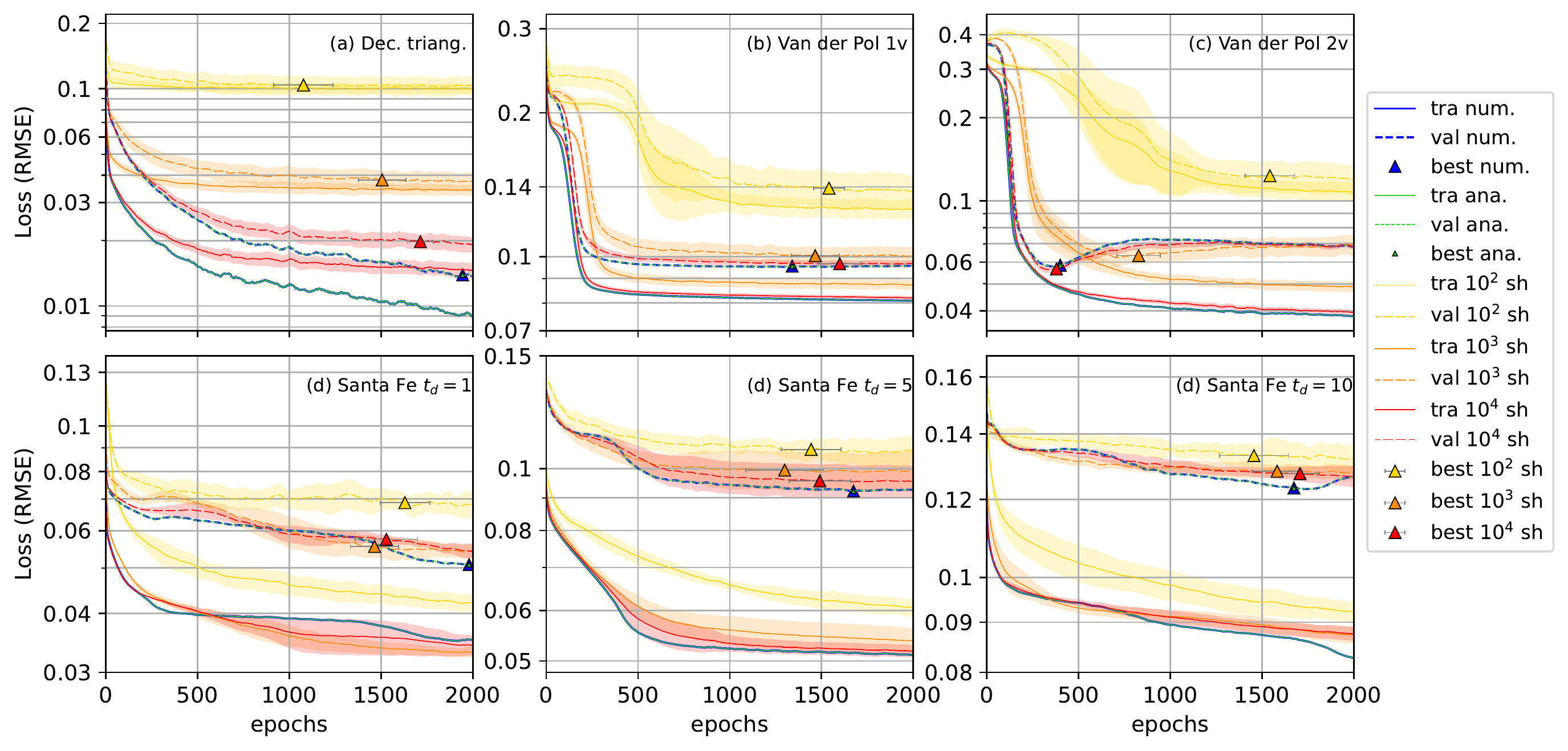}
\caption{Training curves for each of the three datasets, showing how the value of the RMSE varies with the iterations of the parameter optimisation, for training and validation sets, and comparing results with both analytical and numerical gradients. The optimisations are those with lower final validation RMSE for numerical gradients. Optimisations with analytical gradients simulate sampling noise in every circuit evaluation with the number of shots specified in the legend. Lines of optimisations with sampling noise represent the mean of 8 different realisations and shaded regions represent the interval $\pm \sigma$ (standard deviation) around the mean. Triangles indicate the average epoch of minimum validation RMSE  with the corresponding error bars representing the standard error of the mean. The curves are plotted as a moving average of 50 points to ease the visualisation.}
\label{fig:optim_curves}
\end{figure*}

The corresponding datasets and their predictions are shown in Fig. \ref{fig:series}. In addition, Fig. \ref{fig:results_comparison} contains the resulting RMSE distributions of the estimated points with respect to the training, validation and test targets, always considering the last 5 points in each sample. The fourth group of RMSE destributions corresponds to the prediction of the whole test series, not only the last 5 points of each window, by shifting the input windows 5 by 5 points. Fig. \ref{fig:optim_curves} plots the optimisation curves for the training of every dataset. Both Fig. \ref{fig:optim_curves} and Fig. \ref{fig:results_comparison} compare the training and the results when computing numerical gradients and analytical gradients with and without noise.

In terms of final losses, they are identical for numerical and analytical gradients without noise. The univariate output series estimations approximate to the output targets under a reasonable loss because the RMSE is, in the worst case, an order of magnitude lower than the range of output values (from -0.75 to 0.75). Sampling noise causes an effect on the final loss during the training. The loss value cannot be below a threshold that depends on the number of shots, as clearly seen in the optimisation curve of the case (a). This effect is not directly observed in cases (b) and (c) because losses in noiseless optimisations are clearly above this threshold. Analysing Fig. \ref{fig:results_comparison}, the exact evaluation of the RMSE after the optimisation gives for $10^3$ and $10^4$ shots in every dataset region a result that slightly depends on the level of noise. In this sense, conclusions about the training capability of the QRNN with noiseless gradients can be extended to the noisy QRNN, at least for the cases studied. In contrast, $N_{shots} = 10^2$ often represents a limit for the desired precision of predictions.

The convergence curves are also identical for numerical and analytical without noise, thus giving proof of the numerical gradients being very accurate with the chosen step size $\epsilon$. However, they are different for each case, strongly depending on the dataset. The optimiser configuration and the number of iterations were chosen to ensure a good final convergence, and not for minimising the number of epochs. For the six cases, the optimiser reaches RMSE values below 0.1 before 250 epochs with no sampling noise.

However, the optimisation slows down in cases (b) to (f) after the initial sharp drop, and validation loss stops improving from less than 500 epochs onwards. This does not happen in case (a). A possible explanation is the difficulty of finding the correct temporal correlations among input and output series. Moreover, case (c) is a representative case of overfitting, because training and validation curves substantially move away from each other. The effect of sampling noise in this dataset is clear. After the initial drop, the optimisation with $10^2$ and $10^3$ shots is slower than in the ideal emulation, because gradients are less accurate and the optimiser cannot advance so fast when evolving the parameters. This effect is almost not present when evaluating with $10^4$ shots.

The number of epochs to find the best validation loss also depends on the dataset (see markers in Fig. \ref{fig:optim_curves}). In (a), as the optimisation slowly continues converging, it is close to the limit, but very low values are achieved after a few epochs. Only sampling noise can avoid a high accuracy. In (b), the results show no meaningful differences after 500 epochs. In fact, the number of iterations until the solution fluctuates a lot because the curve is almost flat. The noise of $10^2$ and $10^3$ weakly affects the final result, looking at both the curves and the box plots. In (c), the conclusions are the same for $10^4$ shots and analytical, the number of epochs increases with the level of sampling noise (inverse of number of shots). In the noisy emulations, the RMSE values tend to be higher because of the noise. In any case, the final validation is always very close to the noiseless one. Training with dataset (d) reveals plateaus in all cases following the initial drop. Thus, the average best validation point does not vary significantly, regardless of the noise level.

\subsection{Comparison of QRNN with classical models}
We have repeated the forecasting task using classical RNN, LSTM and GRU models, in order to address the capabilities of the QRNN model in comparison with classical common approaches. For a fair comparison, the data distribution and the optimisation procedure are the same as for the variational quantum method, using Adam with learning rate 0.001 and default parameters and sample shuffle in each epoch. Moreover, we set the neural network's hidden size, so that the number of trainable weights is as close as possible to the number of trainable parameters in QRNN. We use \texttt{pytorch.nn} recurrent models plus a linear layer, \texttt{nn.Linear}, for these trainings \cite{paszke2019pytorchimperativestylehighperformance}. The total number of weights is displayed in Table \ref{tab:cml_details}.

\begin{table}[t]
    \centering
\begin{tabular}{l|rrr|rrr|rrr|}
\cline{2-10}
    & \multicolumn{3}{c|}{cRNN}                                                                    & \multicolumn{3}{c|}{LSTM}                                                                    & \multicolumn{3}{c|}{GRU}                                                                     \\ \cline{2-10} 
    & \multicolumn{1}{l|}{$s_{in}$} & \multicolumn{1}{c|}{$s_h$} & \multicolumn{1}{c|}{$N_\theta$} & \multicolumn{1}{l|}{$s_{in}$} & \multicolumn{1}{c|}{$s_h$} & \multicolumn{1}{c|}{$N_\theta$} & \multicolumn{1}{l|}{$s_{in}$} & \multicolumn{1}{c|}{$s_h$} & \multicolumn{1}{c|}{$N_\theta$} \\ \cline{2-10} 
(a) & \multicolumn{1}{r|}{1}        & \multicolumn{1}{r|}{3}     & 22                              & \multicolumn{1}{r|}{1}        & \multicolumn{1}{r|}{1}     & 18                              & \multicolumn{1}{r|}{1}        & \multicolumn{1}{r|}{2}     & 33                              \\ \cline{2-10} 
(b) & \multicolumn{1}{r|}{1}        & \multicolumn{1}{r|}{4}     & 33                              & \multicolumn{1}{r|}{1}        & \multicolumn{1}{r|}{2}     & 43                              & \multicolumn{1}{r|}{1}        & \multicolumn{1}{r|}{2}     & 33                              \\ \cline{2-10} 
(c) & \multicolumn{1}{r|}{2}        & \multicolumn{1}{r|}{6}     & 67                              & \multicolumn{1}{r|}{2}        & \multicolumn{1}{r|}{2}     & 51                              & \multicolumn{1}{r|}{2}        & \multicolumn{1}{r|}{3}     & 67                              \\ \cline{2-10} 
(d) & \multicolumn{1}{r|}{1}        & \multicolumn{1}{r|}{4}     & 33                              & \multicolumn{1}{r|}{1}        & \multicolumn{1}{r|}{2}     & 43                              & \multicolumn{1}{r|}{1}        & \multicolumn{1}{r|}{2}     & 33                              \\ \cline{2-10} 
\end{tabular}
    \caption{Hyperparameters of classical neural networks: input size $s_{in}$, hidden size $s_h$ and the corresponding number of weights $N_\theta$. The output size  $s_{out}$ is consistently set to 1, and the number of layers $L$is fixed at 1, indicating single-layered stacked neural networks.}
    \label{tab:cml_details}
\end{table}

Classical models usually provide better results in the training 
 and validation series, but their performance strongly depends on the dataset and the random parameter initialisation. However, in all the cases, except for (d) with delays 1 and 5, QRNN distribution and classical distributions overlap in both training and validation RMSE. In dataset (a), noiseless QRNN shows a better performance. In the test series, the noiseless quantum model shows less variation compared to the training and validation series than the classical models in most cases, especially for the Santa Fe task with the greater delay. These results are of great interest, since QRNN demonstrates the capacity of generalisation and the ability to forecast over a significant number of future time steps.

\subsection{Comparison of QRNN with a QRC model}

  \begin{figure}
     \centering
     \includegraphics[scale=0.45]{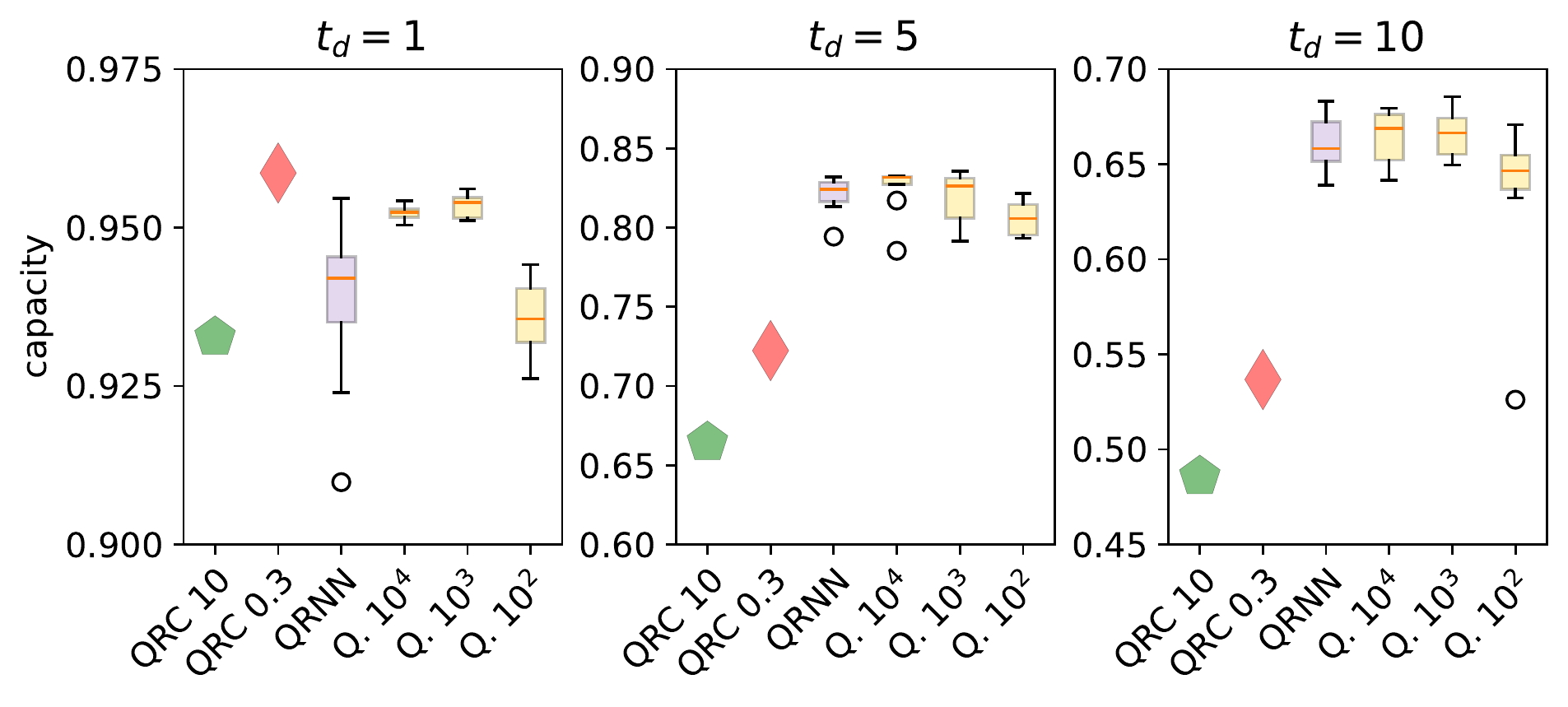}
     \caption{Comparison of results between the study in \cite{mujal_time-series_2023} and the QRNN. The first two points are the capacities in the test series from \cite{mujal_time-series_2023} (Fig. 5). These points correspond to measurement strength $g=0.3$ and $g=10$ respectively. The box plots illustrate the distribution of capacities in the full test region with 10 different trainings of the noiseless QRNN and 8 trainings. These are initialized with the parameters yielding the best RMSE validation among the 10 noiseless optimisations. The labels `Q. $N_{shots}$' denote the results with noisy optimisation and analytical gradients.}
     \label{fig:santafe_compar}
 \end{figure}
 
The problem of extracting information from the quantum circuit was previously addressed in \cite{mujal_time-series_2023}, in the context of Quantum Reservoir Computing. They propose weak measurements, a type of measurement that perturbs the quantum state but does not collapse or destroy it, as a resource for iteratively exchanging information with the quantum circuit and keeping quantum memory simultaneously. The memory capacity of the reservoir is evaluated for different delays, $t_d$, of the output series with respect to the input. That capacity, $C$, is measured by the squared Pearson correlation coefficient, which relates the system outputs and the targets. This metric is scale-invariant.

 In Fig. \ref{fig:santafe_compar} we compare the metric $C$ between our predictions in the complete test region and their predictions in the test region. Both models show similar performances and our method achieves better accuracy for larger delays. However, variational quantum neural networks are harder to train than reservoir computing models. Thus, there is a trade-off between achieving higher accuracy and having faster training. In any case, these results also demonstrate the capabilities of the QRNN as part of the toolkit of quantum algorithms for complex time series prediction.

\section{Discussion \label{sec:conclusions}}
We have used an operator-sum representation for the formulation of the hidden states and the outputs in a Quantum Recurrent Neural Network with intermediate measurements, and we derived its first- and second-order partial derivatives with respect to its trainable parameters. With the density matrix formulation, it is possible to directly perform an ideal emulation of the parameterised quantum circuit that makes up the network and we showed the results for a machine learning task with three different datasets. The analytical gradients permit to train networks with noisy outputs coming from quantum circuits.

The QRNN was presented in previous works \cite{bausch_recurrent_2020,takaki_learning_2021,siemaszko_rapid_2023,bondarenko_learning_2023,li_quantum_2023,sun_quantum-discrete-map-based_2023}. Some of them delve into the inner structure of such quantum circuits, but there was not an explicit formulation to see how quantum states are operated through the quantum circuit. With our work, we contribute to a better understanding of how information is processed inside the network, and we compare it with the classical RNN. Moreover, the given formulas allow a direct implementation of a code for emulation, available through the link in the Data Availability Section.

The operator-sum representation provides various remarkable advantages related to other possible methods. First of all, it avoids computing the complete density matrix of the $n$-qubits quantum state, whose number of elements scales as $\mathcal{O}(4^n)$, by considering the ansatz quantum operator $W$ as a 4-rank tensor.
Secondly, density matrix formulation allows the computation of exact expectation values in circuits with intermediate measurements, which is not feasible with state vector emulation for cases of interest, where series are made up of many points. 

Consequently, the gradients can be computed with high precision in a classical computer. We have seen that parameter optimisations with numerical and analytical gradients and ideal exact emulation converge to identical values. 
That being said, if the precision of the variables is lost, analytical forms become necessary. 
By simulating Gaussian sampling noise, we see that Adam can converge to loss values very close to the noiseless evaluations, in a comparable number of epochs. Therefore, our results point in the direction of using quantum circuits with noisy outputs as predictors of multivariate time series, provided that the optimiser for training supports random variations in the loss function.
The great difference between both methods is the computational cost. Analytical forms require $2T$ circuit evaluations for a gradient component, being $T$ the number of time steps (circuit blocks inside a single circuit).
Meanwhile, a forward finite difference approximation simultaneously shifts the parameter in all circuit blocks, needing only one extra circuit evaluation per gradient component.
Similar conclusions should apply to the Hessian calculation, which requires a $T^2$ factor of circuit evaluations to be computed analytically, but it is usually approximated due to its high computational cost in most cases. 

In this context, the idea of distributed quantum computing, where several Quantum Processing Units work simultaneously, gathers strength. Despite the number of circuit evaluations needed, the QRNN model shows a significant feature: it does not require a lot of qubits. For the hardware-efficient ansatz and the datasets we studied, an ideal quantum processor of 5 qubits would be enough to run the circuits.
In this respect, future work should assess a possible trade-off between the use of quantum computers and the use of analytical gradients with the Parameter Shift Rule, as well as improvements on the ansatz architectures for a better trainability.

In comparison with other models, both classical and quantum, the performance of the QRNN is satisfactory, achieving similar accuracies and prediction capabilities. In that sense, QRNN can be part of the toolkit of Machine Learning algorithms for multivariate series prediction. As a direction for future work, to assess their actual performance when deployed on quantum hardware, we propose developing specific routines for emulating hardware noise and examining its impact on trainability and prediction accuracy.

The results are promising for potential use in real cases. To this end, the model hyperparameters optimisation to enhance the power of the neural network, the adaptability to the quantum hardware limitations, research on circuit structures that capture correlations behind complex datasets and the adaptability of different optimisation techniques are starting points towards a generalised application of Quantum Recurrent Neural Networks to multivariate time series prediction.

\section*{Data availability statement \label{sec:code}}
The data that support the findings of this study are openly available at the following URL/DOI: \url{https://doi.org/10.5281/zenodo.12755194} (Source Code and Training Data).

\section*{Acknowledgments}
We thank the CESGA Quantum Computing researchers for their feedback and the stimulating intellectual environment they provide.
We thank Constantino Rodríguez Ramos, especially for theoretical insights. We also thank P. Mujal and R. Zambrini for the feedback and for sharing their numerical results with us.
This work was supported by Axencia Galega de Innovación through the Grant Agreement ``Despregamento dunha infraestructura baseada en tecnoloxías cuánticas da información que permita impulsar a I+D+I en Galicia'' within the program FEDER Galicia 2014-2020. This work was partially supported by the Galician Government under Grant ED431B 2024/44.
A. Gómez, D. Faílde and M. M. Juane were supported by MICIN through the European Union NextGenerationEU recovery plan (PRTR-C17.I1), and by the Galician Regional Government through the “Planes Complementarios de I+D+I con las Comunidades Autónomas” in Quantum Communication. J. D. Viqueira was supported by Axencia Galega de Innovación (Xunta de Galicia) through the ``Programa de axudas á etapa predoutoral''.
Simulations on this work were performed using Galicia Supercomputing Center (CESGA) FinisTerrae III supercomputer with financing from the Programa Operativo Plurirregional de España 2014-2020 of ERDF, ICTS-2019-02-CESGA-3, and the Qmio quantum infrastructure, with financing from the European Union, through the Programa Operativo Galicia 2014-2020 of ERDF$\_$REACT EU, as part of the European Union’s response to the COVID-19 pandemic.

\section*{Author contributions}
J. D. V., M. M. J. and A. G. conceived the problem.
J. D.V. developed the mathematical model and programmed the emulator.
D.F. reviewed the mathematical model.
All the authors contributed to analyse the data and review the manuscript.

\section*{}
\noindent
\fbox{%
    \parbox{0.98\columnwidth}{%
        \footnotesize This is the Accepted Manuscript version of an article accepted for publication in \textit{Machine Learning: Science and Technology}. IOP Publishing Ltd is not responsible for any errors or omissions in this version of the manuscript or any version derived from it. This Accepted Manuscript is published under a CC BY license. The Version of Record is available online at \href{https://iopscience.iop.org/article/10.1088/2632-2153/ad9431}{10.1088/2632-2153/ad9431}.
    }%
}

\vspace{1em}


\bibliography{dm_emulation_qrnn}

\clearpage

\appendix 
\onecolumngrid

\section{A further explanation on tensor notation \label{sec:tensor_notation}}
In this paper, the mathematical formulation contains tensors of ranks 0 to 5. A main character plus a set of indices identify the tensors. This notation is valuable (i) to have more compact formulas by decreasing the bra-ket notation and (ii) to show the expression to implement in a classical emulation code.

5-rank tensors are represented by an upper shift letter that contains a lower index in parenthesis, indicating the time step (circuit block number). Let $\mathcal{H}_E$ and $\mathcal{H}_M$ be the finite-dimensional complex Hilbert spaces where the state vectors of registers E and M live, respectively, and $\mathcal{H}$ be the Hilbert space of the $n$ qubits. The equivalence for a quantum operator in terms of Hilbert space basis vectors $\{\ket{a},\ket{b},\bra{c},\bra{d} \}$ is

\begin{equation}
    U_{(t)} = \sum_{a,b,c,d} U^{ab}_{(t) cd} \ket{a \otimes b} \bra{c \otimes d}, \: \ket{a} \in \mathcal{H}_E, \bra{c} \in \mathcal{H}_E^*,  \ket{b} \in \mathcal{H}_M, \bra{d} \in \mathcal{H}_M^*,
\end{equation}
where $U_{(t)} \in L(\mathcal{H})$ represents the quantum operator at time $t$, and $U^{ab}_{(t) cd} \in \mathbb{C}$ denotes the numerical value of the corresponding component, that is, the tensor component with index values $a,b,c,d$. Here, $L(\mathcal{H})$ stands for the space of bounded linear operators. The tensor product is represented in such a way that $\ket{a \otimes b} = \ket{a} \otimes \ket{b}$. A density matrix is represented with the same form, represented with the greek character $\rho$.

4-rank tensors are equivalent to 5-rank ones, but omitting the time label $(t)$. 3-rank tensors could appear in different forms, depending on the indices positioning.

2-rank tensors also depend on the index positioning. A lowercase letter with two indices above represents the tensor component of a quantum state living in the $\mathcal{H}_E \otimes \mathcal{H}_M$ space, which satisfies the following equation:

\begin{equation}
    \ket{v} = \sum_{a,b} v^{ab} \ket{a \otimes b}, \: \ket{a} \in \mathcal{H}_E, \ket{b} \in \mathcal{H}_M,
\end{equation}
while the Hermitian conjugate is
\begin{equation}
    \bra{v} = \sum_{a,b} (v^\dagger)_{ij}  \bra{i \otimes j} = \sum_{a,b} (v^{ij})^*  \bra{i \otimes j}, \: \bra{i} \in \mathcal{H}_E^*, \bra{j} \in \mathcal{H}_M^*.
\end{equation}
A symbol with one index above and one below (optionally with a time label, $(t)$) represents an operator or a density matrix living in one of the Hilbert spaces,
\begin{equation}
    U = \sum_{a,c} U^{a}_{c} \ket{a} \bra{c}, \: \ket{a} \in \mathcal{H}_R , \bra{c} \in \mathcal{H}_R^*, R \in \{E,M\}.
\end{equation}

1-rank tensors are actually (state)vectors living in only one of the (dual) Hilbert spaces, $\mathcal{H}_R$ ($\mathcal{H}_R^*$),
\begin{equation}
    \ket{v} = \sum_a v^a \ket{a} \text{ or } w_i = b \bra{i}, \: \ket{a} \in \mathcal{H}_R, R \in \{E,M\}.
\end{equation}
\begin{equation}
    \bra{v} = \sum_a v_a \bra{a} \text{ or } w_i = b \bra{i}, \: \bra{a} \in \mathcal{H}_R^*, R \in \{E,M\}.
\end{equation}

0-rank tensors are represented by a lowercase Latin symbol without any indices, and they are scalars.

We show the equivalences of operations in Eq. \eqref{eq:equivalences}. Indices can appear as single values or pairs, depending on whether the corresponding Hilbert space is $\mathcal{H} \in \{\mathcal{H}_E, \mathcal{H}_M \}$ or $\mathcal{H}_E \otimes \mathcal{H}_M$.
\begin{equation}
    \begin{array}{cll}
        \text{(unitary operation) } &\left( U \ket{v} \right)^i                  &= U^{i}_{k} \: v^{k} \\
        \text{(unitary product) }   &\left( U \cdot V \right)^{i}_{k}            &= U^{i}_{m} \: V^{m}_{k}\\
        \text{(vectors tensor product ) } &\left( \ket{v} \otimes \ket{w} \right)^{ij} &= v^i \: w^j\\
        \text{(vectors outer product) }  &\left( \ket{v} \bra{w} \right)^{i}_{j}      &= v^{i} \: w_{j}^*\\
        \text{(matrix tensor product) } &\left( U \otimes V \right)^{ij}_{kl}        &= U^i_{k} \: V^{j}_{l}\\
        \text{(hermitian conjugate) }   &\left( U^\dagger \right)^{i}_{k}            &= U^{k*}_{i}\\
        \text{(trace)}              &\text{Tr } \rho                             &= \rho^{ik}_{jl} \delta^{jl}_{ik}\\
        \text{(partial trace A) }   &\left( \text{Tr}_A \: \rho \right)^{k}_{l}  &= \rho^{ik}_{il} =  \rho^{ik}_{jl} \: \delta^{j}_{i} \\
        \text{(partial trace B) }   &\left( \text{Tr}_B \: \rho \right)^{i}_{j}  &= \rho^{ik}_{jk} =  \rho^{ik}_{jl} \: \delta^{l}_{k}
    \end{array}
    \label{eq:equivalences}
\end{equation}
Note that the tensor product operation does not commute. Therefore, the writing order does affect the final result. We use the Einstein summation convention where indices that appear twice (up and down) in a term imply the summation of that term over all the values of the index. The symbol $\delta$ is the Kronecker delta and allows to raise and lower indices, as shown in partial trace operations. In order to avoid ambiguities or to highlight the operation, operators, quantum states, and density matrices are defined in the text and explicit summation operators are shown when necessary. 

In Section \ref{subsec:qrnn_memory_output}, summation runs over $N_E$ for indices identifying register E (first index of a pair or index of the first element in a tensor product) and $N_M$ for indices identifying register M (second index or index of the second element).

In the case of Jacobian (gradient) and Hessian elements (Section \ref{sec:gradients}), they can be also thought of as tensors, and the convention is: upper indices for rows, lower indices for columns; first index of a pair for circuit block and second index of a pair for the parameter number.

\section{Derivation of the Parameter Shift Rule for a QRNN circuit \label{sec:psr}}

Analytical partial derivatives of expectation values from circuits based on parameterised rotations are computed as \cite{takaki_learning_2021}
\begin{equation}
    \partial_i \braket{O} = \partial_i \text{Tr} \left[ U \rho_{(t)} U^\dagger O\right] = \text{Tr} \left[ \partial_i U \: \rho_{(t)} \: U^\dagger \: O \right] + \text{Tr} \left[ U \: \rho_{(t)} \: \partial_i U^\dagger \: O \right] + \text{Tr} \left[ U \: \partial_i \rho_{(t)} \: U^\dagger \: O \right],
    \label{eq:derivative_o}
\end{equation}
where the first two terms can be computed with the well-known PSR, by shifting the parameter $\theta_i$ at circuit block $t$,
\begin{equation}
\begin{split}
    \text{Tr} \left[ \partial_i U \: \rho_{(t)} \: U^\dagger \: O \right] + \text{Tr} \left[ U \: \rho_{(t)} \: \partial_i U^\dagger \: O \right] = \\
    \frac{1}{2} \left\{  \text{Tr} \left[ U \left(\theta_i + \frac{\pi}{2} \right) \: \rho_{(t)} \: U^\dagger \left( \theta_i + \frac{\pi}{2} \right) \: O \right] - \text{Tr} \left[ U \left(\theta_i - \frac{\pi}{2} \right) \: \rho_{(t)} \: U^\dagger \left( \theta_i - \frac{\pi}{2} \right) \: O \right] \right\},
\end{split}
\label{eq:psr_simple}
\end{equation}
as in \cite{schuld_evaluating_2019,mitarai_quantum_2018}. For the latter term, 
\begin{equation}
    \text{Tr} \left[ U \: \partial_i \rho_{(t)} \: U^\dagger \: O \right] = \text{Tr} \left[ U \: \partial_i \left( \rho_0 \otimes \rho_{M \: (t-1)} \right) \: U^\dagger \: O \right] = \text{Tr} \left[ U \:  \left( \rho_0 \otimes \text{Tr}_E \left[ \partial_i \left( U \rho_{(t-1)} U^\dagger \right) \right] \right) \: U^\dagger \: O \right],
\end{equation}
where $\rho_0 = \left( \ket{0} \bra{0} \right)^{\otimes n_E}$. By applying the chain rule, we get again three terms,
\begin{equation}
    \partial_i \left( U \rho_{(t-1)} U^\dagger \right) = \partial_i U \rho_{(t-1)} U^\dagger + U \rho_{(t-1)} \partial_i U^\dagger + U \partial_i \rho_{(t-1)} U^\dagger.
\end{equation}
The contribution from the first two is calculated by shifting the parameter $\theta_i$ twice at block $t-1$, as in Eq. \eqref{eq:psr_simple}, while the third one depends on $t-2$ block. Then, the equation is recursive. Each block at $t_k<t$ contributes summing with two terms corresponding to the shifts $+\pi/2$ and $-\pi/2$. The recursion ends at $t_k = 0$, then
\begin{equation}
    \partial_i \braket{O}_{(t)} = \sum_{r=0}^t \frac{1}{2} \left( \left. \braket{O}_{(t)}^+ \right|_{ri} - \left. \braket{O}_{(t)}^- \right|_{ri} \right),
    \label{eq:psr_app}
\end{equation}
where $\left. \braket{O}_{(t)}^\chi \right|_{ri}$ is the expectation value at time $t$ after a $\chi$-shift of the parameter $\theta_i$ at time $t_k$, with $\chi \in \{ +\pi/2, -\pi/2\}$. The shift is represented uniquely by the sign for simplicity. This is the equation in the main text, where we have indicated all the shifts positioning only after the brackets.

For the Hessian, we take every term from Eq. \eqref{eq:psr_app}, $\left. \braket{O}_{(t)}^\chi \right|_{ri}$, and apply again the PSR,
\begin{equation}
    \partial_j \left. \braket{O}_{(t)}^\chi \right|_{ri} = \\
    \text{Tr} \left. \left( \partial_j U' \: \rho_{(t)} \: U'^\dagger  \: O \right)^\chi \right|_{ri} + \text{Tr} \left. \left( U' \: \rho_{(t)} \: \partial_i U'^\dagger \: O \right)^\chi \right|_{ri} + \text{Tr} \left. \left( U' \: \partial_i \rho_{(t)} \: U'^\dagger \: O \right)^\chi \right|_{ri} ,
\end{equation}
being $U' = U \left( \theta_i + \chi \right)$. The subsequent derivation is now analogue to the one for the first-order PSR. From Eq.\eqref{eq:psr_app}, now
\begin{equation}
\begin{split}
    \partial_i \partial_j \braket{O}_{(t)} = 
    \frac{1}{4} \sum_r^t \sum_s^t \left( \left. \braket{O}^{++}_{(t)} \right|^{ri}_{sj} + \left. \braket{O}^{--}_{(t)} \right|^{ri}_{sj}  - \left. \braket{O}^{+-}_{(t)} \right|^{ri}_{sj} - \left. \braket{O}^{-+}_{(t)} \right|^{ri}_{sj} \right),
\end{split}
\label{eq:psr2_app}
\end{equation}
where $\left. \braket{O}^{\chi \lambda}_{(t)} \right|^{ri}_{sj}$ is the expectation value at time $t$ after a shift $\chi$ in the parameter $\theta_i$ at block $r$ and a shift $\lambda$ in the parameter $\theta_j$ at block $s$. Then, we have 4 different circuits per term in sums. For the case $i=j$, $\left. \braket{O}^{\chi \lambda}_{(t)} \right|^{ri}_{sj}$ is invariant under $r \leftrightarrow s$ exchange, then, we do not need to expand the full sum over $s$ in Eq. \eqref{eq:psr2_app}, but only up to $r$. Moreover, when $i=j$ and $r=s$, equal-sign shifts become a single $\pi$-shift and contrary-sign shifts cancel. Then,
\begin{equation}
\begin{split}
    \partial_i^2 \braket{O}_{(t)} = \frac{1}{2} \sum_r^t \sum_s^{r-1} \left( \left. \braket{O}^{++}_{(t)} \right|^{ri}_{si} + \left. \braket{O}^{--}_{(t)} \right|^{ri}_{si} \right) 
    + \frac{1}{2} \sum_r^t \left( \left. \braket{O}^{++}_{(t)} \right|^{ri}_{ri} + \left. \braket{O}^{--}_{(t)} \right|^{ri}_{ri} \right) - \\
    - \frac{1}{2} \sum_r^t \sum_s^{r-1} \left( \left. \braket{O}_{(t)}^{+-} \right|^{ri}_{si} + \left. \braket{O}_{(t)}^{-+} \right|^{ri}_{si} \right) - \frac{1}{2} \sum_r^t \braket{O}_{(t)},
\end{split}
\end{equation}
which is equivalent to the equation in the main text. Since all the possible combinations of $r$ and $s$ are computed, the Hessian is symmetric under $i \leftrightarrow j$ exchange.

The number of different circuits to execute, needed to compute the Hessian of the $T$ observables $\braket{O}_{(t)}$ with $t$ from 0 to $T-1$, is then
\begin{equation}
    4 \times \frac{N_\theta ( N_\theta -1 )}{2} T^2 + 2 \times N_\theta \frac{T(T+1)}{2} + 2 \times N_\theta \frac{T(T-1)}{2} + 1 = 2N_\theta^2 T^2 + 1.
\end{equation}

\section{Machine Learning task details \label{sec:details}}

We have accomplished the machine learning of the three datasets by a classical machine emulating the ideal PQC by an algorithm based on the formulas derived in Section \ref{sec:information}.
The emulation has been simply performed by matrix calculations with \textit{NumPy}, leveraging the reductions in operations provided by the formulation. 
For parameter optimisation, we employed the Adam optimiser in \texttt{qml.AdamOptimizer}, from the open-source library \textit{Pennylane} \cite{bergholm2022pennylane}. 
A DASK client was used for gradient components parallelisation \cite{dask2016}.

The optimiser starts at a randomly generated point in the parameter landscape (except the bias $b$, which is initialised to 0), and we execute 10 different optimisations with different random initialisations. 
Maximum number of iterations is set to 2000.

The three different datasets were generated as explained in the following paragraphs.

Case (a) is a triangular dimmed signal described by the equation
\begin{equation}
    s(t) = A \: e^{-\mu t} \: g (P,t),
\end{equation}
where $g (P,t)$ is a triangular signal of period $P$. Input series is directly got from this equation, while the reference output series is the same series several steps forward in time, i.e.,
\begin{equation}
    y(t) = s(t+t_d) .
\end{equation}

Case (b) is a forced Van der Pol signal, described by the differential equation
\begin{equation}
    \frac{d s^2}{dt^2} - \mu (1-s^2) \frac{ds}{dt} + s = A \: \sin (\omega t),
\end{equation}
where the right-hand side is a sinusoidal perturbation. The input series is
\begin{equation}
    x(t) = c \: s(t),
\end{equation}
while the label is
\begin{equation}
    y(t) = c \: s(t+t_d).
\end{equation}

Case (c) consists of two Van der Pol signals without perturbation, each of them described by the equation
\begin{equation}
    \frac{d s_i^2}{dt^2} - \mu_i (1-s_i^2) \frac{ds_i}{dt} + s_i = 0.
\end{equation}
The input series are given by
\begin{equation}
    x_i(t) = c_i \: s_i(t),
\end{equation}
while the label series is
\begin{equation}
    y(t) = d_0 x_0(t+t_0) + d_1 x_1 (t+t_1).
\end{equation}

All the series are 1000 points, filling times from $t=0$ to 100. Note that in this Appendix, the value $t$ represents the physical time, but not the time step (point) in the time series. The parameters of the dataset generators are summarised in Eq. \eqref{eq:gen_param}.

\begin{equation}
    \begin{array}{crlrlrlrlrlrlrlrl}
     \text{(a)}  &A    &= 0.75 &\mu    &= 0.02 & T     &= 5 & t_d      &= 12   & & & &\\ 
     \text{(b)}  &A    &= 1    &c      &= 0.25 & \mu   &= 2 & \omega   &= 5    & t_d   &= 15& & &\\ 
     \text{(c)}  &c_0  &= 0.25 &d_0    &= 1    & \mu_0 &= 2 & t_0      &= 5    &c_1    &= 0.25  & d_1  &= 0.1  &\mu_1  &= 1 &t_1   &= 18
    \end{array}
    \label{eq:gen_param}
\end{equation}

Dataset (d) was generated from the original Santa Fe laser series with the formula
\begin{equation}
    0.75 \frac{(\bm{x}_0-\bar{x})}{\text{max}(\text{abs} (\bm{x}_0))} ,
\end{equation}
where $\bm{x}_0$ denotes the original series and $\bar{x}$ denotes the average value.

\end{document}